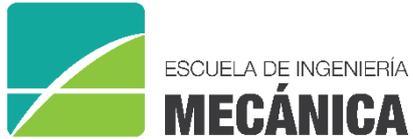 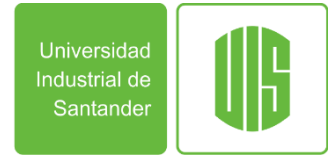

PREPRINT PAPER

# DAMAGE DETECTION IN A UNIDIMENSIONAL TRUSS USING THE FIREFLY OPTIMIZATION ALGORITHM AND FINITE ELEMENTS

Heller Guillermo Sánchez Acevedo
Camilo Andrés Manrique Escobar
Octavio Andrés González-Estrada

A009
Bucaramanga 2018

Research Group on Energy and Environment – GIEMA
School of Mechanical Engineering
Universidad Industrial de Santander

**Abstract**: In this paper, we investigate the damage detection of structures seen as an optimization problem, using modal characterization to evaluate the dynamic response of the structure given a damage model. We implemented the firefly optimization algorithm with a simple numerical damage model to assess the performance of the method and its advantages for structural health monitoring (SHM). We show some implementation details and discuss the obtained results for a benchmark problem.

**Correspondence**: agonzale@uis.edu.co

Research Group on Energy and Environment – GIEMA
School of Mechanical Engineering
Universidad Industrial de Santander
Ciudad Universitaria
Bucaramanga, Colombia

email: giema@uis.edu.co   http://giema.uis.edu.co

# Damage detection in a unidimensional truss using the firefly optimization algorithm and finite elements

# Detección de daños en una armadura unidimensional por medio del algoritmo de optimización de la luciérnaga y elementos finitos

Heller Guillermo Sánchez Acevedo[1], Camilo Andrés Manrique Escobar[1], Octavio Andrés González-Estrada[2]

[1] GIEMA, School of Mechanical Engineering, Universidad Industrial de Santander, Ciudad Universitaria, Bucaramanga, camilo.manrique@correo.uis.edu.co, hsanchez@uis.edu.co
[2] GIC, School of Mechanical Engineering, Universidad Industrial de Santander, Ciudad Universitaria, Bucaramanga, Colombia, Orcid: 0000-0002-2778-3389, Email: agonzale@uis.edu.co

## Abstract
In this paper, we investigate the damage detection of structures seen as an optimization problem, using modal characterization to evaluate the dynamic response of the structure given a damage model. We implemented the firefly optimization algorithm with a simple numerical damage model to assess the performance of the method and its advantages for structural health monitoring (SHM). We show some implementation details and discuss the obtained results for a benchmark problem.

## Resumen
El presente trabajo tiene por objeto investigar la detección de daños en estructuras como un problema de optimización, utilizando la caracterización modal para evaluar la respuesta dinámica de la estructura ante un modelo de daño. Se llevó a cabo la implementación del algoritmo de optimización de la luciérnaga, tomando como caso de estudio un modelo numérico de daño sencillo, para ver los alcances del método propuesto y sus ventajas dentro del campo de monitoreo estructural (SHM). Se indican detalles de la implementación y los resultados obtenidos con este planteamiento para un problema propuesto.

## Palabras clave
Algoritmo de la luciérnaga, optimización, elementos finitos, análisis modal, monitoreo estructural.

## 1. Introduction

The importance of early detection of damage has become more important during the last decades in industry and academy, going through visual inspection to vibrational analysis [1]–[3]. Structural health monitoring (SHM) systems have been implemented widely in bridges such as the Great Belt Bridge in Denmark, the Confederation Bridge in Canada, among others [4]. The identification of the location and the depth of cracks in elements of the structure have received considerable attention in the Structural Health Monitoring (SHM) field.

Damage detection methods can be classified into two categories: those based on identification with dynamic data, and methods based on identification with static data (deformations and stiffness matrices), the latter having less information available for analysis, as well as difficulty to find damage in components of structures whose contribution to total deformation is low [5].

Damage detection in structures through modal analysis is one of the most used methods including dynamic data of structures. This based on the fact that, deterioration of the condition of a structure or element is linked to the loss of stiffness and damping, affecting the dynamic properties of the system. The variation presented in modal parameters is an indicator of the magnitude and localization of damage [6], [7]. SHM using dynamic response can be classified into two groups according to the implementation [8]: experimental methods based on non-destructive techniques and numerical methods based on FEA (Finite Element Analysis) [2], where the use of this analytical tool with previous experimental validation is a great asset in the process [9]. The last one has particularly led to the development of intelligent structures or systems, which are capable of detecting damage online and quantifying the degree of severity of damage.

In the literature, there are different methods for the detection of damage based on the analysis of the dynamic parameters due to changes in the rigidity of the system [6], [10]–[12]. In [10], neural networks combined with fuzzy pattern recognition were used to do an online categorization of the health state of a bridge among four categories (healthy, little damage, moderate damage, and significant damage) through FEA computations of natural frequencies. This algorithm, although its satisfactory results in categorization, is not able to estimate damage quantity or location.

In [11], genetic algorithms were considered to solve the problem of the detection of damage in structures and machine elements, which was addressed as an optimization problem.

In seminal works, binary codified genetic algorithms were used, the objective function was based on the residual forces vector [12]. Such approach has the disadvantage of requiring full modal forms, which in practice is currently not feasible due to technical and economic reasons.

Later works proposed to detect the damage dividing the process into two stages to define its location and magnitude [13]. In the first stage, a set of elements possibly damaged was determined through a methodology of locating elements with damage based on energy. In the second stage, the damage is quantified using a micro-genetic algorithm, which performs an optimization process, where the optimal combination of damaged elements and damage extensions is sought to minimize a target function based on natural frequencies and modal forms.

In [14], a modified genetic algorithm for the detection of structural damage was used. The algorithm considers a chromosome representation defined with real numbers and an objective function based on changes in natural frequencies and modal forms. Subsequently, it restarts the individuals who present a minimum difference in the objective function to define the new population. This type of coding is highly applicable to solve the problem of damage detection, in view of the fact that the number and position of the damage elements are not known *a priori*.

The work of [15] analyzes in detail the rigidity matrix of cracked beams and the non-linearity linked to the opening employing fracture mechanics. Fault detection is performed by comparing natural frequencies of the models. They conclude that plate or brick elements in FEA analysis are not necessary for SHM techniques for this type of structures. However, further work is required in the area since the developed algorithm satisfactorily detects the location of the damage, but gives an estimate of damage minor than the actual value.

In [16], the failure is modeled by fracture mechanics and using FEA the dynamic mode shapes are determined, allowing to associate the modal stiffness with the work of the internal forces and the displacement field. Then, the virtual work is calculated as a scalar product of the mode shapes. This allows computing the Modal Assurance Criterion (MAC), knowing that virtual work is the square root of it. Based on this, a failure indicator for a life-time and service-life estimation is proposed.

A framework with FEA updating simulations for probabilistic lifetime estimation for wind energy converter structures is developed by [17]. A detailed model of the structure is created using the FEA software ANSYS to compute the eigenvalues of the system, then a simplified one, with less computational cost, is validated, obtaining nearly perfect matches in the eigenvalues. The assumption of the reliability of the highly detailed FEA model is corroborated in [18] finding a 1.3% deviation with respect to measured data. This is used to develop a tool for online SHM. The eigenvalues are obtained by Operational Modal Analysis (OMA) of the real structure and through MAC-matrix diagonal values, an optimization problem with an objective function of the similarity between real-time simulations allows the identification of damage. A damage catalog with patterns is created *a priori* with FEA to allow rapid assessment of damage.

In the present work, we propose the use of a metaheuristic optimization algorithm, called Firefly Algorithm (FA), for the detection of structural damage through modal characterization. The swarm algorithm considers variations in the dynamic response of the structure, given a simple damage model, which is characterized by a decrease in the elastic properties of the damaged elements. The modal characterization is done by means of a finite element (FE) numerical model. In the next section, we present the problem statement and the definition of the optimization problem using FA. Then, numerical results are presented for a benchmark problem, where different damage configurations are considered. We perform the analysis of the different optimization parameters and their impact in the proposed FA algorithm. Finally, we present the most relevant conclusions.

## 2. Methodology

### 2.1. Problem Statement

Consider the problem of free vibration without damping. The mathematical model that defines the equation of motion for the system with one degree of freedom can be written as:

$$m\ddot{u}(t) + ku(t) = 0, \qquad (1)$$

where *m* is the mass, *k* is the elastic constant and *u* is the displacement solution of the system, which depends on the time. The general solution of (1) is:

$$x = A\sin(\omega t + B), \omega = \sqrt{k/m}, \qquad (2)$$

Where *A* and *B* are real constants and *ω* represents the natural frequency of the system. Equation (2) can be generalized for systems of several degrees of freedom as:

$$[\mathbf{M}][\ddot{\mathbf{u}}(t)] + [\mathbf{K}][\mathbf{u}(t)] = 0, \qquad (3)$$

where **M** is the mass matrix, **K** is the stiffness matrix of the system, and **u** is the displacement field. The solution of (3) is not unique and for a system of $n$ degrees of freedom there are $n$ solutions $\mathbf{u}_i$ or mode shapes, each one associated with a natural frequency of the system $\omega_i$.

## 2.2. Finite element formulation and modal analysis

For a continuous system, it is possible to find the solution by means of a finite element discretization where, in equation (3), **u** represents the nodal displacements $\mathbf{u}^h$ and matrices **M** and **K** are constructed from the matrices evaluated in the domain of each element, defined as:

$$\mathbf{M}^e = \int_{\Omega_e} \mathbf{N}^T \rho \mathbf{N} \, d\Omega,$$

$$\mathbf{K}^e = \int_{\Omega_e} \mathbf{B}^T \mathbf{D} \mathbf{B} \, d\Omega, \quad (4)$$

where $\rho$ is the mass per unit volume, **N** are the basis functions used for the finite element approximation and **B** their derivatives.

## 2.3. Firefly Algorithm

The firefly algorithm was proposed by [19], inspired by the behavior of fireflies. It is a metaheuristic optimization algorithm for swarm intelligence. This algorithm offers advantages of operation when searching in extensive solution spaces since it does not have a starting point and it avoids falling into local optimum, improving its performance in the global space [20]. Figure 1 presents the pseudocode proposed by Yang.

```
FIREFLY ALGORITHM
Objective function f(x), x = [x_1, ..., x_d]^T.
Generate an initial population of n fireflies x_i (i = 1, 2, ..., n).
Light intensity I_i at x_i, is determined by f(x_i).
Define light absorption coefficient γ.
while (t < MaxGeneration),
    for i = 1 : n (all n fireflies)
        for j = 1 : n (all n fireflies) (inner loop)
            if (I_i < I_j)
                Move firefly i towards j.
            end if
            Vary attractiveness with distance r via e^{-γr^2}.
            Evaluate new solutions and update light intensity.
        end for j
    end for i
    Rank the fireflies and find the current global best g∗.
end while
Postprocess results and visualization.
```

Figure 1. Pseudocode of the FA algorithm.

Fireflies use light to attract other fireflies during mating. In the algorithm, the light intensity can be formulated in such a way that it is associated with the objective function to be optimized. The algorithm generates a determined number of possible solutions within the search field, in which the configurations (fireflies) that give the best response of the objective function will be those that will attract the other configurations that are close. In this way, key points of the solution field are examined in a more efficient way. The control parameters of the FA algorithm are defined as:

$x$: Population of fireflies.
$n$: Size of the population, total number of fireflies.
$I_i$: Light intensity of the firefly $i$.
$MaxGeneration$: Maximum number of generations for fireflies.
$\alpha$: Sets the randomness of the process. It defines the step in the movement of the fireflies.
$\gamma$: Coefficient of light absorption.
$\Delta$: It establishes the reduction of randomness whenever a new generation originates.
$\beta$: Attractiveness. Coefficient of attraction between fireflies, which varies with respect to the light absorption coefficient and the distance between them, given by

$$\beta(r) = \beta_0 e^{-\gamma r^m}, m > 1 \tag{5}$$

where $\beta_0$ is the attractiveness at $r = 0$.

Yang [9] established three fundamental rules governing the algorithm, which determine the behavior pattern of fireflies. These rules include:

- Fireflies are unisex, meaning a firefly will be attracted to another firefly regardless of sex.
- The attractiveness of a firefly is proportional to its brightness, and these two (attractiveness and brightness) are forced to decrease when the distance from another firefly increases. Less bright fireflies will move toward one with greater brightness. If a firefly is not attracted to any other, because there is none with a brightness greater than its own, it will move randomly.
- The brightness of a firefly is obtained evaluating the objective function.

## 2.6. Objective Function

The optimization problem for the detection of structural damage can be defined through a functional that expresses the weighted quadratic difference between the response of the healthy model and the damaged model. The objective function to be optimized, proposed by [1], is:

$$f(x) = \sum_{j=1}^{n_\omega} W_{\omega_j} \left[1 - \left(\frac{\omega_{mj}}{\omega_{aj}}\right)\right]^2 + \sum_{j=1}^{n_\omega}\sum_{i=1}^{s} W_{\phi_{ji}} \left(\phi_{mji} - \phi_{aij}\right), \tag{6}$$

where the quadratic difference is minimized as a function of the eigenvalues $\omega$ and the eigenvectors $\phi$ of the problem under consideration. $W_{\omega_j}$ and $W_{\phi_{ji}}$ are the weighting factors for $\omega$ and $\phi$ respectively, $n_\omega$ represents the number of eigenvalues to be considered and $s$ defines the size of the eigenvector.

In this sense, the objective function will be in charge of the comparison of the dynamic characteristics of the test model and the models of the database that has preconfigurations of possible damages.

## 2.7. Noise

In order to simulate distortion in the signals of the test model, noise was introduced in the values of natural frequencies $\omega$ and modal forms $\phi$. Noise is expressed in terms of small sums or subtractions given by

$$\omega_r = \omega(1 + Rand(-1,1)N_\omega), \tag{7}$$

$$\phi_r = \phi(1 + Rand(-1,1)N_\phi), \tag{8}$$

where $N_i$ represents the percentages of noise to be used in eigenvalues and eigenvectors, randomly distributed.

## 2.8. Algorithm description

For the solution of the problem two programs are used: Ansys APDL and Matlab. First, it is required to evaluate the dynamic response for different damage scenarios. In Matlab, a database containing the dynamic responses for different damage configurations (affected bars and percentage of elasticity reduction) is constructed offline. Ansys APDL code is used to simulate stiffness losses in the bars and to obtain the modal characterization.

With the database already built, the firefly algorithm is implemented. The algorithm generates a given number of fireflies (possible solutions within the database) that explore the solution field. The fireflies are evaluating the objective function and, by means of the intensity, they are grouped in the optimal local of the solution space. The optimization problem proposed by the objective function in (6) compares the test model with different damage models and finds the one that best fits the input data. The general flowchart of the proposed algorithm is shown in
Figure 2.

## 3. Numerical results

To evaluate the performance of the proposed methodology, we implement the problem of a truss subjected to damage in different configurations. Noise is introduced into the model to simulate the effect of problems in signal capture and processing. We modify different optimization parameters to evaluate the response of the FA algorithm.

### 3.1 Truss

The planar truss in Figure 3 is composed of 13 bars of steel A-36, with Young's modulus $E = 2 \times 10^{11}\ Pa$, Poisson's ratio $\nu = 0.3$ and, density $\rho = 7850\ Kg/m^3$. The structure is simply supported on nodes 1 and 5, as shown. The cross-sectional area of the bars is $A = 4 \times 10^{-4}\ m^2$. The total height of the structure is 2.4284 m and the span is 7.3152 m. The problem is to determine if any of the bars of the structure have a failure that could put the operation at risk. For the numerical model of the structure implemented in Ansys we considered LINK1 elements. We solve the problem of free vibration without damping in equation (3) to obtain the eigenvalues and eigenvectors of the system. Initially, in order to set the optimization problem defined by the objective function in (6), the weighting is defined as $W_{\omega j} = 1$ and $W_{\phi ji} = 1$. The configuration parameters of the FA algorithm, as indicated in [10], are taken as $\alpha = 0.2$, $\beta_0 = 1$, $\gamma = 1$, and $\Delta = 0.97$.

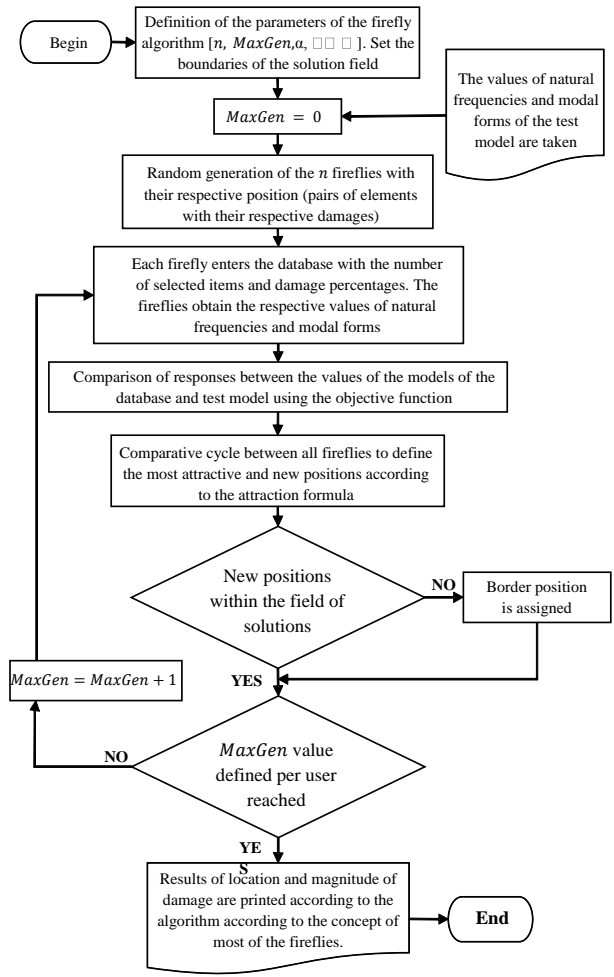

Figure 2. Flowchart of the proposed algorithm.

The database containing different damage configurations is generated offline to aid in the search for damage in the input model. The database contains the responses for damages in one or more bars within a range of [0, 95] in damage percentage, discretized every 5%. For the solution of (3), and to limit the problem size, only the extraction of the first 8 modal forms was considered.

The optimization algorithm FA searches within a precalculated database, which contains the dynamic response of the structure for different faults. We determine the damaged elements and quantify the loss of rigidity for each element by means of the objective function.

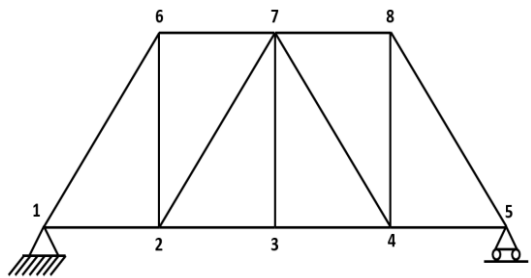

Figure 3. Truss simply supported.

## 3.1 Damage detection

In order to evaluate the response of the proposed algorithm against noise, a set of tests are performed. It is necessary to identify the minimum number of mode shapes required for the algorithm to behave appropriately, even in the presence of noise in the input data. To do that, a set of noise-scenarios are defined following (7), as **N**=[$N_\omega, N_\phi$]: **N1**= [0%, 0%], **N2**= [0.5%, 1%], **N3**= [1%, 3%], **N4**= [2%, 5%], where $N_\omega$ represents the noise added to the natural frequency value and $N_\phi$ is the value added to the mode shapes, for a corresponding input.

Thus, accuracy tests were performed in the proposed algorithm to verify how it responds to the noise, trying to emulate the response to an input signal with distortions. Three attempts were made for each combination of damage configuration and number of modes extracted, and the percentage of effectiveness was obtained.

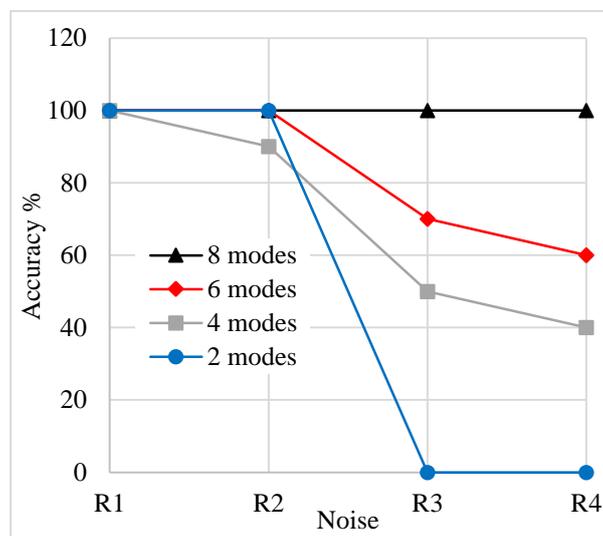

Figure 4. Accuracy of the algorithm versus different configurations of noise in the test model: **N1** = [0%, 0%], **N2** = [0.5%, 1%], **N3** = [1%, 3%], **N4** = [2%, 5%].

Figure 4 shows the effectiveness of the algorithm against noise for different number of modal forms extracted. The results show that when we consider only 2 mode shapes the optimization algorithm is very sensitive to noise and is not able to find the damaged elements in some cases. Increasing the number of mode shapes between 4 and 6 increases the effectiveness of the method by having an enriched search space. For 8 modes, the percentage of effectiveness of the algorithm looking for damaged elements was 100%. Therefore, for subsequent analysis of the performance of the optimization algorithm, we choose to extract 8 mode shapes.

Figure 5 shows the results of the tests performed to identify the sensitivity of the algorithm to the noise in the natural frequency, modes of vibration and both. We evaluate the accuracy to detect the magnitude and location of the damage when a discrete range of noise is added to the sensitivity parameters. The input variables *Damage* and *Position* are generated randomly in order to consider the performance of the algorithm throughout the entire database. Once the noise effect is added, the detection algorithm is executed and the predictions are obtained, which are compared with the original input variables to evaluate the accuracy. This process is replicated ten times to identify an average accuracy behavior given the random nature of the metaheuristic.

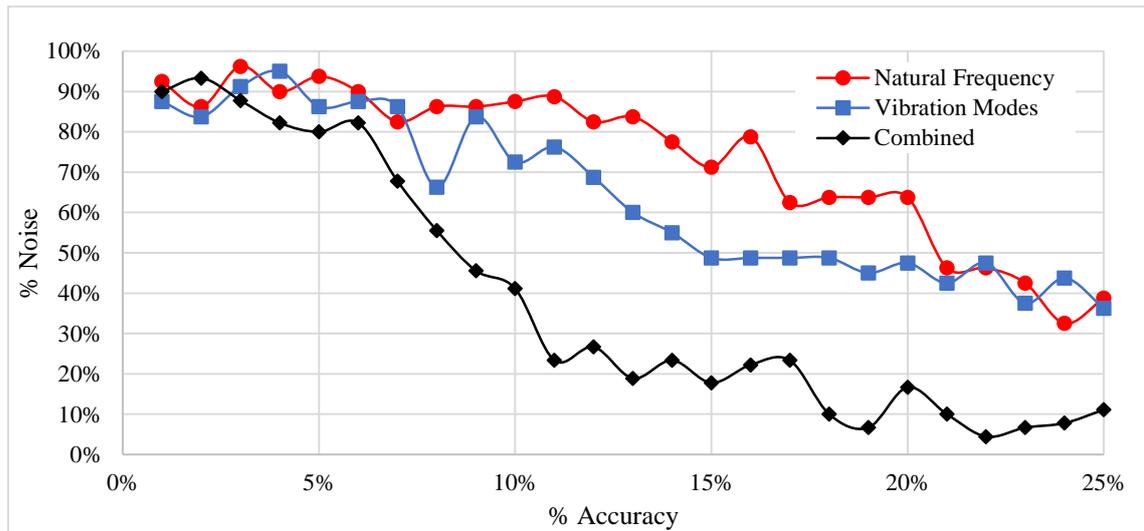

Figure 5. Accuracy of the model versus noise.

It can be observed that an accuracy of 80% can be expected if accelerometers with a ±5% noise in their measurements are employed. The algorithm shows excellent performance in detecting the position of damage given that noise in the eigenvectors turns them into scalar multiples of the original unaffected ones, allowing the algorithm to identify location very accurately. For the magnitude of the damage, the results are shown in Table 1, where red boxes indicate an incorrect result and green ones a correct result, it can be observed for one of the ten data sets, how accuracy in the prediction on the magnitude of the damage decades for increasing values of noise.

| Iteration | 4% | | 7% | | 10% | | 13% | | 16% | | 19% | | 22% | | 25% | |
|---|---|---|---|---|---|---|---|---|---|---|---|---|---|---|---|---|
| 1 | P | D | P | D | P | D | P | D | P | D | P | D | P | D | P | D |
| 2 | P | D | P | D | P | D | P | D | P | D | P | D | P | D | P | D |
| 3 | P | D | P | D | P | D | P | D | P | D | P | D | P | D | P | D |
| 4 | P | D | P | D | P | D | P | D | P | D | P | D | P | D | P | D |
| 5 | P | D | P | D | P | D | P | D | P | D | P | D | P | D | P | D |
| 6 | P | D | P | D | P | D | P | D | P | D | P | D | P | D | P | D |
| 7 | P | D | P | D | P | D | P | D | P | D | P | D | P | D | P | D |
| 8 | P | D | P | D | P | D | P | D | P | D | P | D | P | D | P | D |
| 9 | P | D | P | D | P | D | P | D | P | D | P | D | P | D | P | D |
| 10 | P | D | P | D | P | D | P | D | P | D | P | D | P | D | P | D |

Table 1. Damage performance against noise. P: Location of damage, D: Magnitude of damage.

A more detailed analysis of the magnitude of damage detection is shown in Table 2. For noise values of 16%, the algorithm shows good estimations in the damage magnitude being those near the actual value in the input data.

| | NOISE % | | | | | | | | | | | | | | | |
|---|---|---|---|---|---|---|---|---|---|---|---|---|---|---|---|---|
| | 4% | | 7% | | 10% | | 13% | | 16% | | 19% | | 22% | | 25% | |
| Iter | In | Out | In | Out | In | Out | In | Out | In | Out | In | Out | In | Out | In | Out |
| 1 | 30-85 | 30-85 | 45-80 | 45-80 | 60-25 | 45-20 | 55-50 | 45-40 | 95-70 | 95-70 | 35-60 | 15-50 | 85-75 | 20-25 | 90-85 | 45-75 |

| 2 | 45-75 | 45-75 | 55-80 | 55-80 | 85-35 | 85-35 | 85-80 | 70-75 | 30-65 | 30-50 | 50-15 | 40-10 | 85-90 | 40-85 | 95-60 | 90-35 |
| 3 | 15-35 | 15-35 | 10-50 | 10-50 | 65-25 | 60-25 | 70-90 | 45-80 | 55-25 | 55-25 | 65-80 | 65-80 | 85-85 | 85-75 | 95-65 | 70-20 |
| 4 | 20-75 | 20-75 | 60-10 | 60-10 | 55-15 | 45-10 | 25-85 | 15-85 | 20-80 | 5-70 | 60-30 | 35-30 | 95-65 | 90-30 | 10-60 | 10-60 |
| 5 | 30-20 | 30-20 | 5-15 | 65-15 | 45-10 | 45-10 | 25-35 | 5-25 | 15-65 | 20-55 | 45-90 | 10-75 | 50-15 | 30-10 | 95-50 | 30-30 |
| 6 | 45-45 | 45-45 | 55-15 | 55-15 | 85-25 | 85-25 | 45-65 | 45-65 | 15-25 | 10-25 | 55-40 | 20-35 | 40-45 | 25-35 | 95-95 | 50-80 |
| 7 | 60-75 | 60- 75 | 50-65 | 50-65 | 75-25 | 75-25 | 30-40 | 30-35 | 15-80 | 15-80 | 45-90 | 45-80 | 95-50 | 65-40 | 95-50 | 90-45 |
| 8 | 90-60 | 90- 60 | 85-75 | 85-70 | 85-25 | 85-25 | 75-60 | 70-60 | 45-80 | 10-75 | 95-50 | 55-25 | 25-55 | 40-85 | 85-15 | 80-15 |
| 9 | 95-20 | 95- 20 | 40-10 | 35-10 | 75-45 | 75-45 | 35-50 | 30-45 | 75-65 | 60-55 | 50-55 | 30-45 | 45-50 | 45-50 | 80-40 | 30-30 |
| 10 | 70-40 | 70- 40 | 20-10 | 20-10 | 85-85 | 85-85 | 25-50 | 25-50 | 30-35 | 10-10 | 70-75 | 20-25 | 55-85 | 25-20 | 95-70 | 85-55 |

Table 2. Damage prediction performance under noise conditions. In: input data, Out: output data.

For this reason, an accuracy analysis of the algorithm to identify the location of the damage under the effect of the noise in the input data is executed in the same fashion as the previously shown, but considering only the location of the damage to compute the accuracy value.

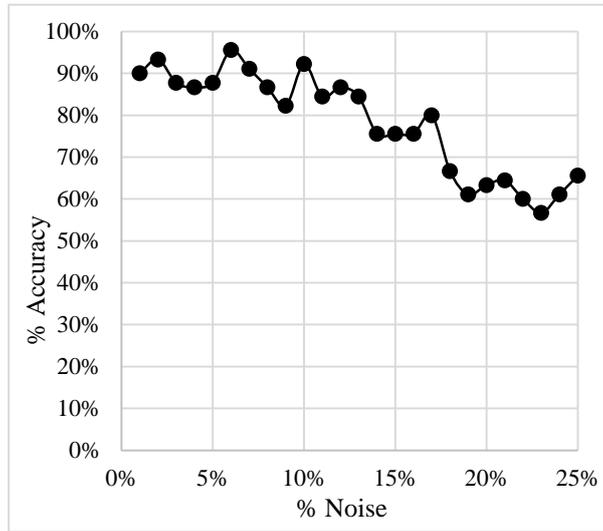

Figure 6. Accuracy of the model against noise considering only the damage location.

From this test, it can be observed in Figure 6 that the algorithm presents high accuracy in the detection of damage location under the presence of noise in the input data. Thus, the algorithm can be considered as a tool for locating the damage and estimating the percentage of damage of the affected elements.

### 3.2 Analysis of metaheuristic parameters

In order to identify the most relevant parameters in the performance of the Firefly Algorithm, a $2^k$ experimental design is proposed according to [21] The parameters $n$ =[25 40], *MaxGeneration*= [2500 5000], $\gamma$= [0.21 1], with the dependent variable $f$ = value of the objective function, as defined in (6).

From Table 3, we can conclude that the most relevant factors in the performance of the Firefly Algorithm are $n$ and *MaxGeneration*, since they have a P value lower than 0.05 and, based on this, it is inferred that they are statistically significant.

|  | Effect | Coef. | Se | T value | P value |
|---|---|---|---|---|---|
| **Constant** |  | 0.141 | 0.007 | 19.89 | 0 |
| n | -0.06 | -0.03 | 0.007 | -4.24 | 0 |
| γ | -0.009 | -0.004 | 0.007 | -0.64 | 0.526 |
| MaxGeneration | -0.031 | -0.015 | 0.007 | -2.19 | 0.032 |
| n γ | 0.002 | 0.001 | 0.007 | 0.19 | 0.848 |
| n MaxGeneration | 0.011 | 0.005 | 0.007 | 0.84 | 0.404 |
| γ MaxGeneration | -0.027 | -0.013 | 0.007 | -1.9 | 0.062 |
| n MaxGeneration γ | 0.02 | 0.01 | 0.007 | 1.42 | 0.159 |

Table 1. Parameter effect analysis.

In Figure 7 the standardized effects of each of the factors and their respective interactions are presented. Each of the effects of the 3 parameters considered and their interactions are in descending order. The dashed red line indicates a critical value with 95% confidence, where the values farthest to the right of the latter will be relevant. It is observed that the effects of the parameters *n* and *MaxGeneration* are the most significant and, additionally, the parameter *n* is the most relevant of all.

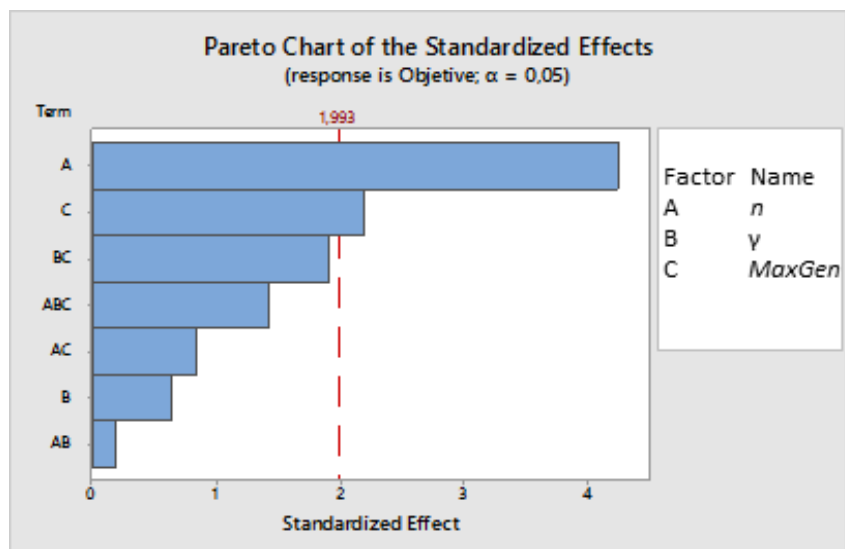

Figure 7. Pareto chart of the standardized effects.

## 4. Conclusions

In the present work, a methodology was developed that addresses the problem of the detection of damage in structures as an optimization problem. ANSYS finite element software was used for the dynamic characterization of the structures. For the solution of the optimization problem, the Firefly Algorithm was chosen as the tool to determine the damage condition in the models.

The methodology was based on the fact that damage affects the stiffness of the system, which results in a change in its dynamic response. A database was constructed offline with eigenvalues and eigenvectors precalculated for different types of damage. The algorithm evaluates the dynamic response of the test model and compares it with the database to determine the magnitude and location of damage.

The results show that, for various damage configurations and a moderate noise percentage (cases R1, R2, R3), the algorithm is able to detect the affected elements or regions and quantify the damage percentage. When there is a severe noise condition (case R4) it is necessary to consider a higher number of mode shapes (6 or 8) to obtain a reliable response from the algorithm. This behavior is due to the fact that, when working with few vibration modes, the algorithm does not have enough information on the nodal displacements to determine the location and magnitude of the damage. We can conclude that the number of shape modes to be extracted should be 8 or higher.

A more exhaustive analysis of the effect of noise for this condition is performed by running tests where the natural frequency, shape modes and both are affected. From these analyses, it is concluded that the noise in the shape modes has a greater negative effect on the accuracy of the results. For tests with noise affecting both natural frequency and shape modes it is observed that the algorithm presents an acceptable accuracy, as low as 68% within a range of 7% of noise, this considering the estimation of location and magnitude of the damage. When considering the accuracy of only the location of damage, a much better performance is obtained, getting for 13% noise an accuracy of 84%. Hence, the algorithm can be considered as a tool for damage location and estimation of its magnitude.

From the metaheuristic parameters analysis, it is concluded that the most relevant parameter is $n$, size of the population, total number of fireflies. Then, computational resources must be focused on increasing this value to obtain better performance of the algorithm.

## 5. References

[1]   K. Moslem and R. Nafaspour, "Structural Damage Detection by Genetic Algorithms," *AIAA J.*, vol. 40, no. 7, pp. 1395–1401, May 2012.
[2]   O. A. González-Estrada, J. Leal Enciso, and J. D. Reyes Herrera, "Análisis de integridad estructural de tuberías de material compuesto para el transporte de hidrocarburos por elementos finitos," *Rev. UIS Ing.*, vol. 15, no. 2, pp. 105–116, Jan. 2016, doi:10.18273/revuin.v15n2-2016009.
[3]   A. M. Agredo Chávez, S. J. Sarmiento Nova, Á. Viviescas Jaimes, and A. Viviescas Jaimes, "Evaluación de la rigidez a flexión de puentes de viga-losa en concreto presforzado a partir de pruebas de carga. Caso de estudio: puente La Parroquia, vía La Renta - San Vicente de Chucurí," *Rev. UIS Ing.*, vol. 15, no. 2, pp. 145–159, 2016.
[4]   J. M. Ko and Y. Q. Ni, "Technology developments in structural health monitoring of large-scale bridges," *Eng. Struct.*, vol. 27, pp. 1715–1725, Oct. 2005, doi:10.1016/j.engstruct.2005.02.021.
[5]   F. Bakhtiari-Nejad, A. Rahai, and A. Esfandiari, "A structural damage detection method using static noisy data," *Eng. Struct.*, vol. 27, pp. 1784–1793, Oct. 2005, doi:10.1016/j.engstruct.2005.04.019.
[6]   H. G. Sanchez *et al.*, "Application of Vibration Based Damage Identification Techniques on Metallic Structures," *Adv. Mater. Res.*, vol. 875–877, pp. 875–879, 2014, doi:10.4028/www.scientific.net/AMR.875-877.875.
[7]   H. G. Sánchez Acevedo, J. Uscátegui, and S. Gómez, "Metodología para la detección de fallas en una estructura entramada metálica empleando las técnicas de análisis modal y PSO," *Rev. UIS Ing.*, vol. 16, no. 2, pp. 43–50, 2017.
[8]   Y. Zou, L. Tong, and G. P. Steven, "Vibration-based model-dependent damage (delamination) identification and health monitoring for composite structures — a review," *J. Sound Vib.*, vol. 230, no. 2, pp. 357–378, 2000, doi:10.1006/jsvi.1999.2624.
[9]   C. R. R. Farrar and K. Worden, "An introduction to structural health monitoring," *Philos. Trans. R. Soc. A Math. Phys. Eng. Sci.*, vol. 365, no. 1851, pp. 303–315, 2007, doi:10.1098/rsta.2006.1928.
[10]  M. M. M. Reda Taha and J. Lucero, "Damage identification for structural health monitoring using fuzzy pattern recognition," *Eng. Struct.*, vol. 27, no. 12 SPEC. ISS., pp. 1774–1783, Oct. 2005, doi:10.1016/j.engstruct.2005.04.018.

[11] A. Rytter, "Vibration Based Inspection of Civil Engineering," Aalborg University, 1993.
[12] J. F. Schutte and A. A. Groenwold, "Sizing design of truss structures using particle swarms," *Struct. Multidiscip. Optim.*, vol. 25, no. 4, pp. 261–269, Oct. 2003, doi:10.1007/s00158-003-0316-5.
[13] S. M. Bland and R. K. Kapania, "Damage Identification of Plate Structures Using a Hybrid Genetic-Sensitivity Approach," *AIAA J.*, vol. 43, no. 2, pp. 439–442, May 2005.
[14] J. E. Laier and J. D. V Morales, "Computational Structural Engineering: Proceedings of the International Symposium on Computational Structural Engineering, held in Shanghai, China, June 22--24, 2009," Y. Yuan, J. Cui, and H. A. Mang, Eds. Dordrecht: Springer Netherlands, 2009, pp. 833–839.
[15] M. I. I. Friswell and J. E. T. E. T. Penny, "Crack Modeling for Structural Health Monitoring," *Struct. Heal. Monit. An Int. J.*, vol. 1, no. 2, pp. 139–148, Jun. 2002, doi:10.1177/1475921702001002002.
[16] W. B. Krätzig and Y. S. Petryna, "Structural damage and life-time estimates by nonlinear FE simulation," *Eng. Struct.*, vol. 27, no. 12 SPEC. ISS., pp. 1726–1740, Oct. 2005, doi:10.1016/j.engstruct.2005.04.015.
[17] D. Hartmann, K. Smarsly, and K. Law, "Coupling sensor-based structural health monitoring with finite element model updating for probabilistic lifetime estimation of wind energy converter structures," in *Proceedings of the 8th international workshop on structural health monitoring*, 2011, pp. 13–15.
[18] K. Smarsly, K. Law, and D. Hartmann, "Towards life-cycle management of wind turbines based on structural health monitoring," in *Proceedings of the First International Conference on Performance-based Life-cycle Structural Engineering*, 2012.
[19] X. Yang, *Nature-Inspired Metaheuristic Algorithms*, 1st ed. Frome: Luniver Press, 2008.
[20] X. Yang and X. He, "Firefly algorithm: recent advances and applications," *Int. J. Swarm Intell.*, vol. 1, no. 1, p. 36, Aug. 2013, doi:10.1504/IJSI.2013.055801.
[21] D. Montgomery, *Design and Analysis of Experiments*, 9th ed. Hoboken, NJ: John Wiley & Sons, 2017.